\documentclass[preprint,showpacs,preprintnumbers,amsmath,amssymb]{revtex4}

\usepackage{graphicx}
\usepackage{dcolumn}
\usepackage{bm}

\begin{document}

\preprint{}

\title{Acceptance of fluorescence detectors and its implication in
energy spectrum inference at the highest energies}

\author{Vitor de Souza}
 \email{vitor@astro.iag.usp.br}
 \affiliation{Instituto de Astronomia, Geof\'{\i}sica e Ci\^encias Atmosf\'ericas, Universidade de S\~ao Paulo} 
\author{Gustavo Medina-Tanco}
  \email{gmtanco@gmail.com}
   \affiliation{Instituto de Astronomia, Geof\'{\i}sica e Ci\^encias Atmosf\'ericas, Universidade de S\~ao Paulo} 
\author{Jeferson A. Ortiz}
  \email{jortiz@astro.iag.usp.br}
  \affiliation{Instituto de Astronomia, Geof\'{\i}sica e Ci\^encias Atmosf\'ericas, Universidade de S\~ao Paulo}

\date{\today}
             
\begin{abstract}

Along the years HiRes and AGASA experiments have explored the fluorescence and the
ground array experimental techniques to measure extensive air showers, being both
essential to investigate the ultra-high energy cosmic rays. However, such
Collaborations have published contradictory energy spectra for energies above the
GZK cut-off. In this article, we investigate the acceptance of fluorescence telescopes
to different primary particles at the highest energies. Using CORSIKA and CONEX shower
simulations without and with the new pre-showering scheme, which allows photons to
interact in the Earth magnetic field, we estimate the aperture of the HiRes-I telescope
for gammas, iron nuclei and protons primaries as a function of the number of simulated events and primary
energy. We also investigate the possibility that systematic differences in shower
development for hadrons and gammas could mask or distort vital features of the cosmic
ray energy spectrum at energies above the photo-pion production threshold. The impact of
these effects on the true acceptance of a fluorescence detector is analyzed in the
context of top-down production models.
\end{abstract}


\maketitle

\section{Introduction}
\label{sec:introduction}

Since the very first observations at the beginning of the $20^{\mathrm {th}}$ century,
ultra-high energy cosmic rays have been a priority in astroparticle physics
and an open question.
The flux of particles is one of the most intriguing question to be
answered. Measurements of the energy spectrum of particles allows
us to make assumptions about a great number of astrophysical
models, such as, acceleration mechanisms, magnetic field in the
source and in the medium, propagation mechanisms and anisotropy.

At energies above $10^{19.5}$~eV, the flux of particles is
expected to drop abruptly due to the interaction of the particles
with the microwave background. Such effect, widely known as the
GZK cut-off \cite{bib:g,bib:zk}, imposes sources of cosmic rays in
the neighborhood of approximately 100 Mparsec. Inside this region,
only a few classes of astrophysical objects have been argued to be
able to accelerate particles at these high energies. Moreover, the
energy threshold for pion photoproduction on the microwave
background is $\sim$2$\times$10$^{10}$~GeV, and at
3$\times$10$^{11}$~eV the energy-loss distance is about
20~Mpc~\cite{bib:todor}.

Nevertheless, several experiments have published detections of
cosmic ray particles with energies above the GZK cut-off.
Presently, the AGASA~\cite{bib:agasa} and HiRes~\cite{bib:hires}
Collaborations are the experiments with highest aperture at
energies above $10^{19.5}$~eV. The HiRes Collaboration has firmly
settled the GZK cut-off prediction along the years. On the other
hand, the AGASA Collaboration has argued for the non existence of
the drop in the flux.

Many theoretical works have tried to overcome such scientific
disagreement and production mechanisms have been proposed in order
to explain the excess of high energy particles seen by the AGASA
Collaboration in relation to the expected GZK cut-off. An
explanation of the highest energy cosmic rays is the top-down
scenario which considers the decay of relic particles of the early
universe, i.e., topological defects or super-heavy particles which
may decay into jets of hadrons, mainly pions that subsequently
decay into $\gamma$-rays, electrons and neutrinos. Only a few
percent of the particles produced by these mechanisms are hadrons
which might result in $\gamma$-ray and neutrinos as the principal
component generated in these process.

HiRes and AGASA experiments have explored different experimental
techniques, the fluorescence and the ground array, respectively,
to  measure extensive air showers.  The ground array efficiency is
based on a straightforward detection given by its operational area
which is, in principle, not dependent on the primary particle
type. On the other hand, fluorescence telescopes measure the
longitudinal air shower development by detecting the fluorescence
light emitted along the track of particles in the atmosphere.
Particles with different mass induce showers with distinct
longitudinal evolution in the atmosphere and for this reason the
efficiency of fluorescence telescopes depends on the primary
shower composition.

The HiRes Collaboration states in
reference~\cite{bib:hires:espectro} that the HiRes-I telescope
aperture investigation has been based on simulations of proton and
iron nuclei initiated showers and no mention to gamma-ray showers
has been done so far. In the present work we calculate the HiRes-I
telescope aperture for gamma showers in the energy range from
$10^{19}$~eV to $10^{20.6}$~eV. The corresponding spectrum for the
bottom-up and top-down scenarios are calculated and compared to
the AGASA data. Furthermore, we have studied the determination of
the aperture as a function of the number of simulated showers and
for different simulation codes.

This paper is structured as follows. In Section
\ref{sec:longitudinal} we briefly discuss the longitudinal
development of extensive air showers induced by different primary
particles. In Section \ref{sec:simulation} the shower and detector
simulation schemes are described. The reconstruction procedure and
the analysis, as well, are characterized in Section
\ref{sec:reconstruction}, while the HiRes-I telescope aperture is
depicted in Section \ref{sec:aperture}. An spectrum exposition is
made in Section \ref{sec:spectrumm}. Section \ref{sec:conclusion}
summarizes our results.

\section{Longitudinal development of shower initiated by different primary particles}
\label{sec:longitudinal}

Fluorescence telescopes measure the longitudinal development of
extensive air showers by detecting the fluorescence light emitted
along the track of particles in the atmosphere. Showers initiated
by particles with different mass are known to have different
longitudinal development in the atmosphere.

Since the original proposal of the Gaisser-Hillas
function~\cite{bib:gaisser:hillas}, many
studies~\cite{bib:hires:2,bib:song} have shown that the number of
charged particles (\emph{N}), as a function of the atmospheric
depth (\emph{X}), in hadronic-induced air showers is well
described by a four parameter ($N_{\mathrm{max}}$, $X_0$,
$X_{\mathrm{max}}$ and $\lambda$) function according to equation:

\begin{equation}
N(X) = N_{\mathrm{max}} \left( \frac{X - X_0}{X_{\mathrm{max}} - X_0} \right) ^ \frac{(X_{\mathrm{max}} -
  X_0)}{\lambda} \exp{ \frac{X_{\mathrm{max}} - X}{\lambda} }
\label{GHillas}
\end{equation}

In equation \ref{GHillas} only $N_{\mathrm{max}}$ and
$X_{\mathrm{max}}$ have clear physical meaning in which
$N_{\mathrm{max}}$ is defined as the maximum number of charged
particles in the shower while $X_{\mathrm{max}}$ represents the
depth at which the maximum number of charged particles occurred.
We suggest the reference \cite{bib:pryke:long} which is an
interesting summary about the physical meaning of each parameter.

The HiRes Collaboration~\cite{bib:hires:espectro:2} has shown that
their data are completely independent from the chosen values of
$X_0$ and $\lambda$. Correspondingly, for the HiRes-I analysis they have fixed the values of these parameters to $-60$~g/cm$^2$ and $70$~g/cm$^2$,
respectively.

Both $N_{\mathrm{max}}$ and  $X_{\mathrm{max}}$ have been widely
used to reconstruct measured extensive air showers and to
determine shower primary particle, since the energy and mass
composition can be directly estimated as a function of these
parameters. In addition to the elongation rate
method~\cite{bib:hires:elongation}, the $X_{\mathrm{max}}$
parameter has also been used to distinguish showers initiated by
photons from those initiated by hadrons~\cite{bib:pre:shower:2}.

Not long ago, a new phenomenon has been shown to play a important
rule in the development of gamma showers. Gammas with energies
above $10^{19.0}$~eV can interact with the Earth magnetic field
and produced pairs of electron/positrons which can, by inverse
Compton effect, emit photons. Such photons induce an
electromagnetic cascade before the shower reaches the atmosphere,
i.e., a pre-shower. The geomagnetic field effect shows significant
importance in the development of gamma-initiated air showers. The
overall effect of the pre-shower, concerning the longitudinal
development, is to distort the gamma shower so that they show an
atmospheric development more similar to proton showers. Results
and application of the pre-shower program can be seen in
references
\cite{bib:pre:shower:2,bib:pre:shower:1,bib:pre:shower:3,bib:pre:shower:4}.

\begin{figure}[t]
  \begin{center}
    \includegraphics[width=10cm]{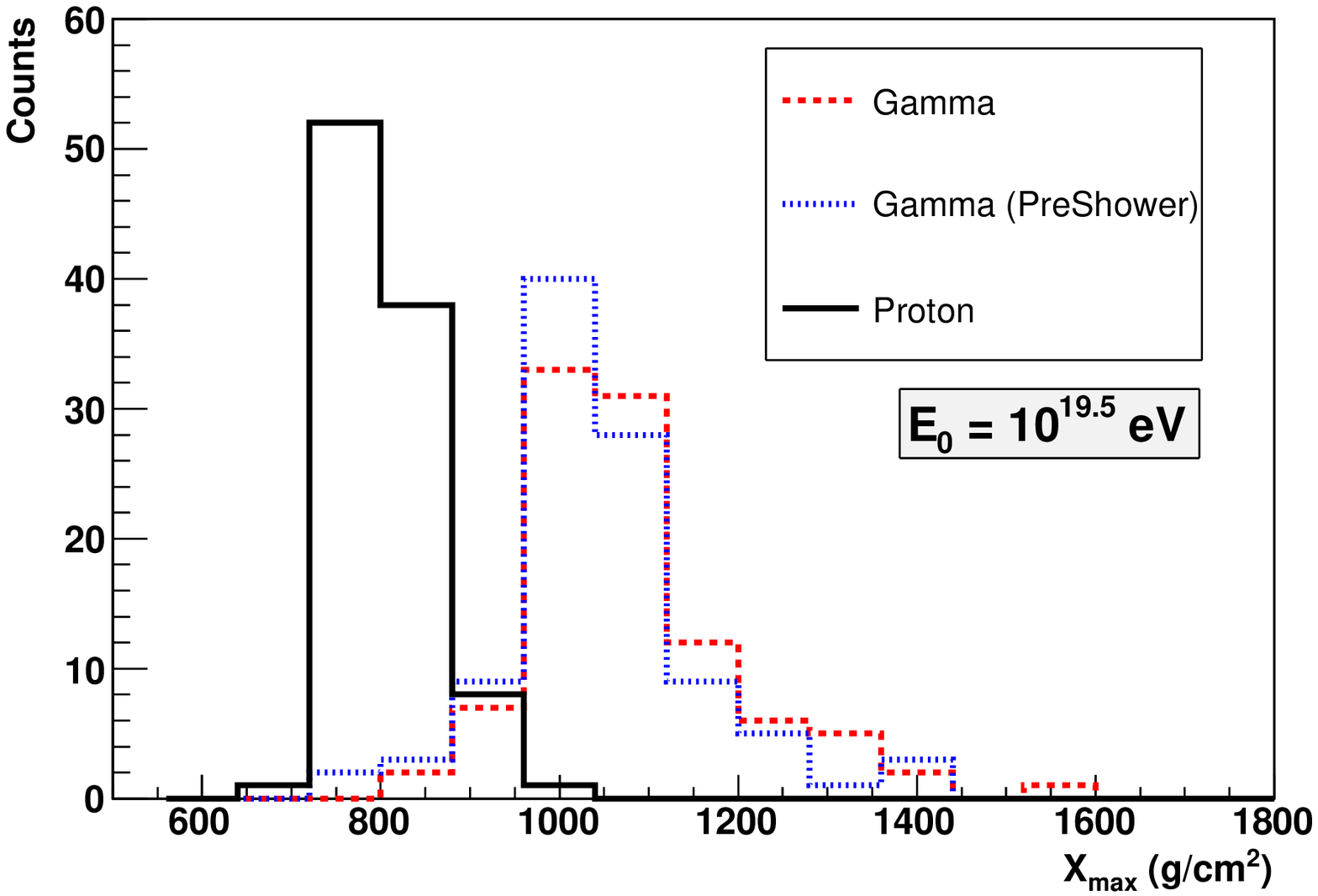}
  \end{center}
   \caption{Distribution of  $X_{\mathrm{max}}$ for showers initiated by
   proton and gamma at $10^{19.5}$ eV. Gamma shower
   were simulated with and without the pre-shower effect. For each
   energy and primary we simulated 100 showers.}
   \label{fig:xmax:19.5}
\end{figure}

\begin{figure}[t]
  \begin{center}
    \includegraphics[width=10cm]{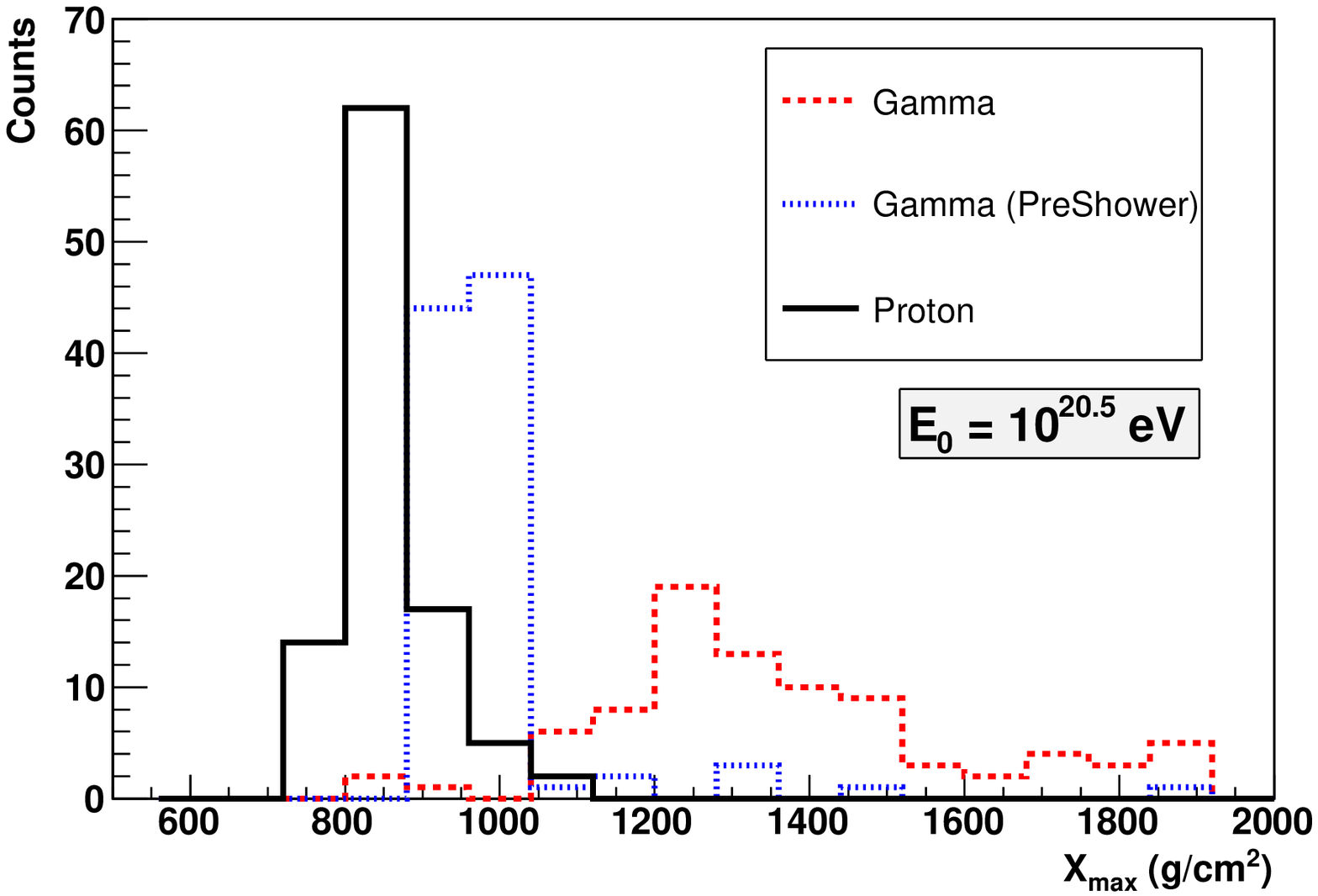}
  \end{center}
   \caption{Distribution of  $X_{\mathrm{max}}$ for showers initiated by
   proton and gamma at $10^{20.5}$ eV. Gamma shower
   were simulated with and without the pre-shower effect. For each
   energy and primary we simulated 100 showers.}
   \label{fig:xmax:20.5}
\end{figure}

Figs.~\ref{fig:xmax:19.5} and \ref{fig:xmax:20.5} show the
$X_{\mathrm{max}}$ distribution of 100 showers originated by
different primary particles. The pre-shower effect is illustrated
by the $X_{\mathrm{max}}$ distribution at $10^{19.5}$ and
$10^{20.5}$ eV. At  $10^{19.5}$ eV the simulation of gamma showers
with and without the pre-shower are almost identical and only a
small reduction of  $X_{\mathrm{max}}$ is seen. At $10^{20.5}$ eV
the separation between the distributions of $X_{\mathrm{max}}$
simulated with and without the pre-shower effect is very clear.

Fig.~\ref{fig:xmax:20.5} also illustrates how the pre-shower
effect approximates the development of a gamma shower to that of a
proton shower. However, even at high energies the
$X_{\mathrm{max}}$ distribution of gamma showers are
distinguishable from proton showers.

The detection of one shower by a fluorescence telescope depends on
the entire development of the shower in the atmosphere. In
particular, for the HiRes-I telescope, it is possible to calculate
the minimum number of particles in a shower as a function of depth
which would trigger the photomultiplier camera. Considering a
background noise of 40 photoelectrons per $\mu s $ in each
photomultiplier and applying a simple trigger condition of signal
higher than four times the noise, it is possible to calculate for
a given geometry the minimum number of particles in the shower
which would trigger the photomultipliers.

\begin{figure}[t]
  \begin{center}
    \includegraphics[width=12cm]{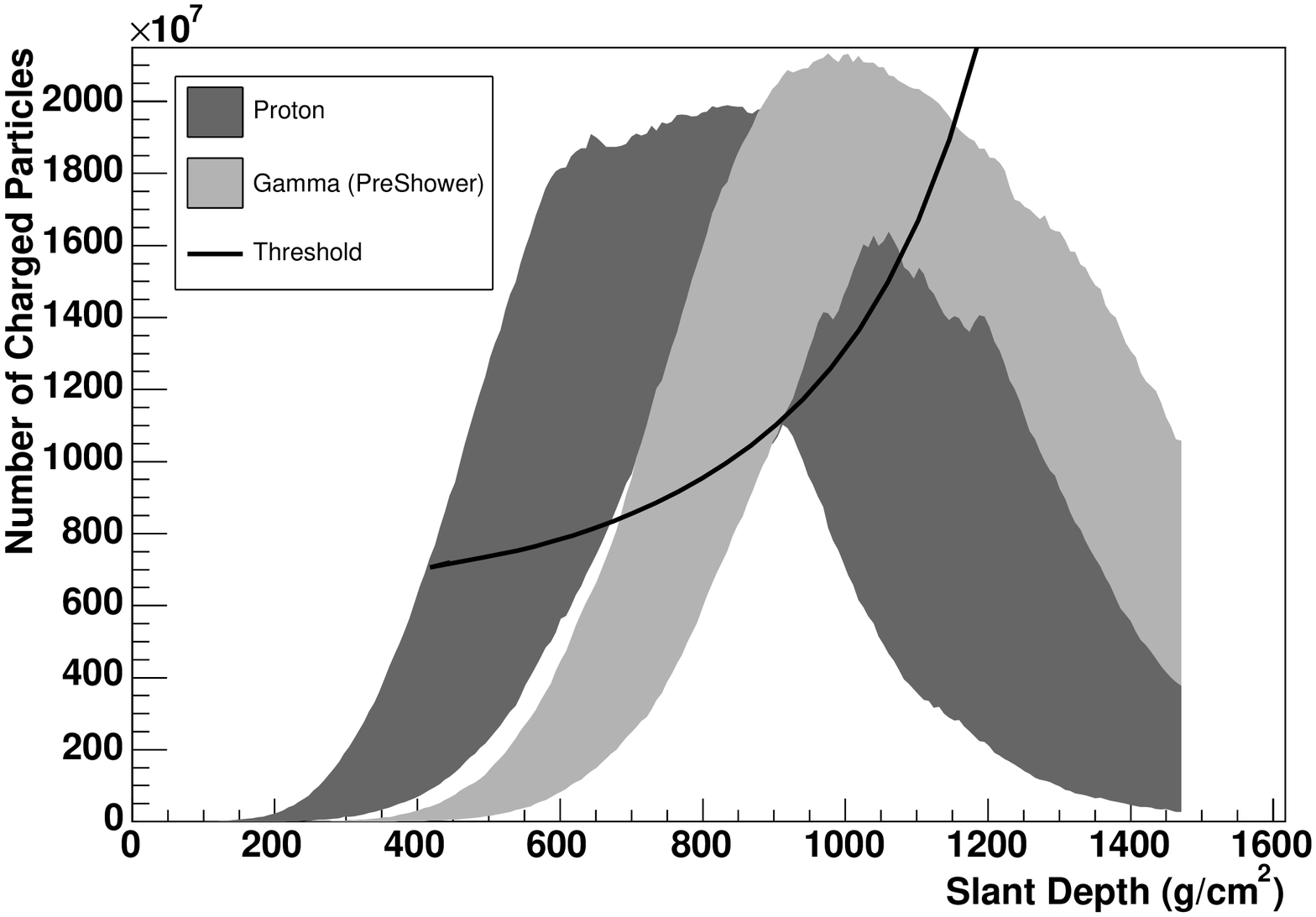}
  \end{center}
   \caption{Distribution of  $X_{\mathrm{max}}$ for showers initiated by
   proton and gamma at $10^{19.5}$ and $10^{20.5}$ eV. Gamma shower
   were simulated with and without the pre-shower effect. For each
   energy and primary we simulated 100 showers.}
   \label{fig:longitudinal}
\end{figure}

Fig.~\ref{fig:longitudinal} illustrates median and one sigma
limits of the the longitudinal development of showers initiated by
proton and gamma primaries (with pre-shower effect). In the same
plot, we show the threshold for a shower landing 25 km away from
the telescope with zenith angle of 45 degrees. It can be seen that
the telescope has a higher probability of triggering under a
proton shower than under a gamma shower.

Besides the trigger conditions, the events used to construct the
spectrum must survive the severe quality cuts imposed by the
reconstruction of geometry and energy. The reconstruction quality
of a shower depends on the size of the seen track and, in
particular, the energy reconstruction error increases markedly if
$X_{\mathrm{max}}$ is not in the field of view of the telescope.

Following the ideas underlined here, we investigate in the
following sections the details involved in the probability of
detection and reconstruction of showers initiated by different
types of primary particles.

\section{Shower and Detector Simulation}
\label{sec:simulation}

We used Monte Carlo simulations to evaluate the aperture of the
HiRes-I telescope. The simulations preformed for this work
followed the same general procedure adopted by the HiRes
Collaboration as published in \cite{bib:hires:espectro} and
\cite{bib:hires:espectro:2}. In these articles, the HiRes
Collaboration has published the aperture of the experiment and all
the important parameters used in the shower and detector
simulation.

Following the HiRes Collaboration prescription, we have performed
a very detailed simulation of the cosmic ray showers and of the
HiRes-I telescope. Showers were simulated using the
CORSIKA~\cite{bib:corsika} program with the
QGSJet~\cite{bib:qgsjet} model for hadronic interactions. The
thinning factor of the simulation was set to $10^{-6}$ and the
longitudinal profile were sampled in steps of 5 g/cm$^2$. Energy
thresholds were set to 0.1 MeV, for electrons and photons, 0.3 GeV
for hadrons and 0.7 GeV for muons. Protons and gamma ray showers
were simulated in the energy interval from $10^{19}$ to
$10^{20.5}$ eV in steps of $0.1$~dex. For each energy and primary
particle we simulated 100 events.

For gamma ray showers, we have simulated two sets: a) with the
pre-shower effect b) without the pre-shower effect. A recent
release of CORSIKA (version 6.2) includes the pre-shower
effect~\cite{bib:pre:shower}.

The probability of interaction of gammas with the geomagnetic
field depends on the relative orientation between the local field
and the momentum of the photons. Therefore we have used in our
simulations the Earth magnetic field for the HiRes site
($112^\circ$ W and $40^\circ$ N) as given by
reference~\cite{bib:mag:field}.

Each CORSIKA shower is used several times to save computational
time by drawing a zenith angle and a core position. For the gamma
showers simulated with the pre-shower effect we only draw a new
core position keeping constant the zenith angle. This is necessary
because the development of the shower depends on the relative
direction of the magnetic field with the primary particle.

For proton and iron nuclei showers, besides using CORSIKA, we have
also simulated a second set of events with the recently released
CONEX program~\cite{bib:conex,bib:conex:0}. This program is based
on a hybrid simulation scheme combining fast numeric solutions of
cascade equations and Monte Carlo calculations. The main
advantages of the hybrid programs is the execution
time~\cite{bib:seneca:corsika}. We have used CONEX in order to: a)
test it in comparison to CORSIKA, regarding specifically the
determination of the aperture of fluorescence telescopes, b) study
the aperture for iron nuclei showers and c) evaluate the possible
biases introduced by the recycling of the same shower several
times, a common approach when using full simulation codes. Since,
CONEX takes on average only 1.5 minutes to simulate one shower in
our current system, we were able to run 5000 different showers for
some energies and compare the result with 100 different CORSIKA
showers used 50 times each. Results are presented in Section
\ref{sec:conex}.

From the longitudinal development of the charged particles in the
atmosphere simulated by CORSIKA or CONEX, it is possible to
calculate the number of fluorescence photons along the shower path
using the fluorescence yield as measured by Kakimoto et
al.~\cite{bib:kakimoto}. Despite the fact that recent
publications~\cite{bib:risse:dedx,bib:nagano:2,bib:bruce:gap} have
shown it is more direct and even more accurate to use the
longitudinal energy deposit of particles we have chosen to use the
longitudinal particles development in order to repeat exactly the
HiRes Collaboration calculations made in
references~\cite{bib:hires:espectro,bib:hires:espectro:2}

We therefore wrote a computer program to simulated the HiRes-I
telescope which propagates the fluorescence photons to the
detector by taking into account scattering and transmission
effects (Rayleigh and Mie) according to \cite{bib:flyseye}.
Atmospheric conditions at the HiRes site have been carefully
measured and the relevant parameters can be found at
\cite{bib:hires:espectro,bib:hires:espectro:2}.

Fluorescence photons are generated taking into account the average
energy deposit as a function of shower age, atmospheric pressure
dependence, and missing energy \cite{bib:song}.


A complete wavelength dependent calculation was done. The
fluorescence spectrum is taken from Bunner et al.
\cite{bib:bunner} and the corresponding normalization applied
according to Kakimoto et al. \cite{bib:kakimoto}. A detailed
simulation of the detector was performed including effective
collection area, mirror reflectivity, filters transmission and
phototube quantum efficiency as given in references
\cite{bib:song:tese,bib:jose:tese}.

\begin{figure}[t]
  \begin{center}
    \includegraphics[width=10cm]{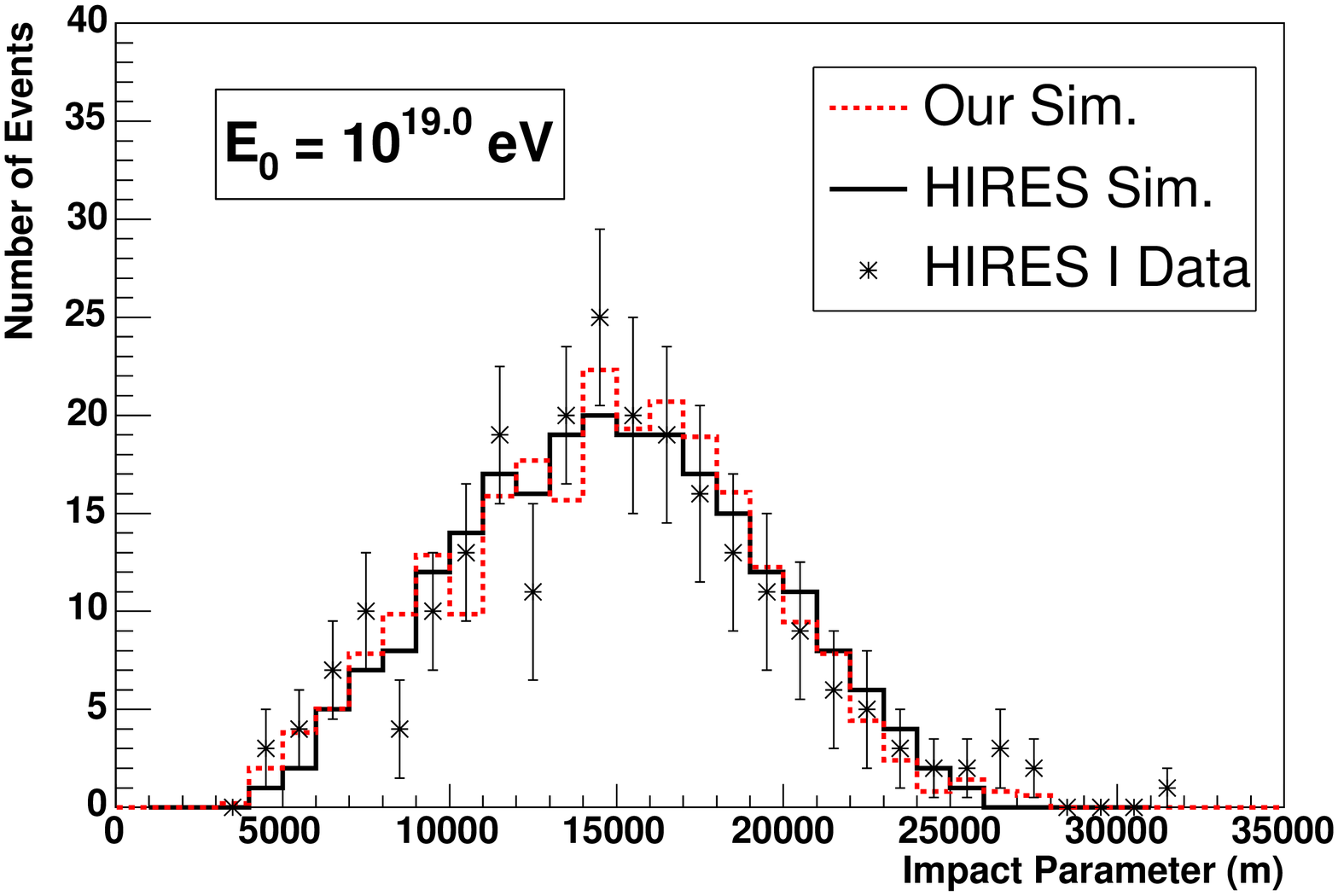}
  \end{center}
   \caption{Distribution of impact parameter for shower initiated with
   energies $10^{19}$ eV. We compare our simulation
   with the HiRes-I data and the Hires Collaboration Simulation. The
   $\chi^2$ between the HiRes data and the HiRes simulation is 1.6 and the
   $\chi^2$ between the HiRes data and our simulation is 1.7. The
   HiRes data and simulation have been extracted from \cite{bib:hires:espectro}.
   }
   \label{fig:rp19}
\end{figure}

\begin{figure}[t]
  \begin{center}
    \includegraphics[width=10cm]{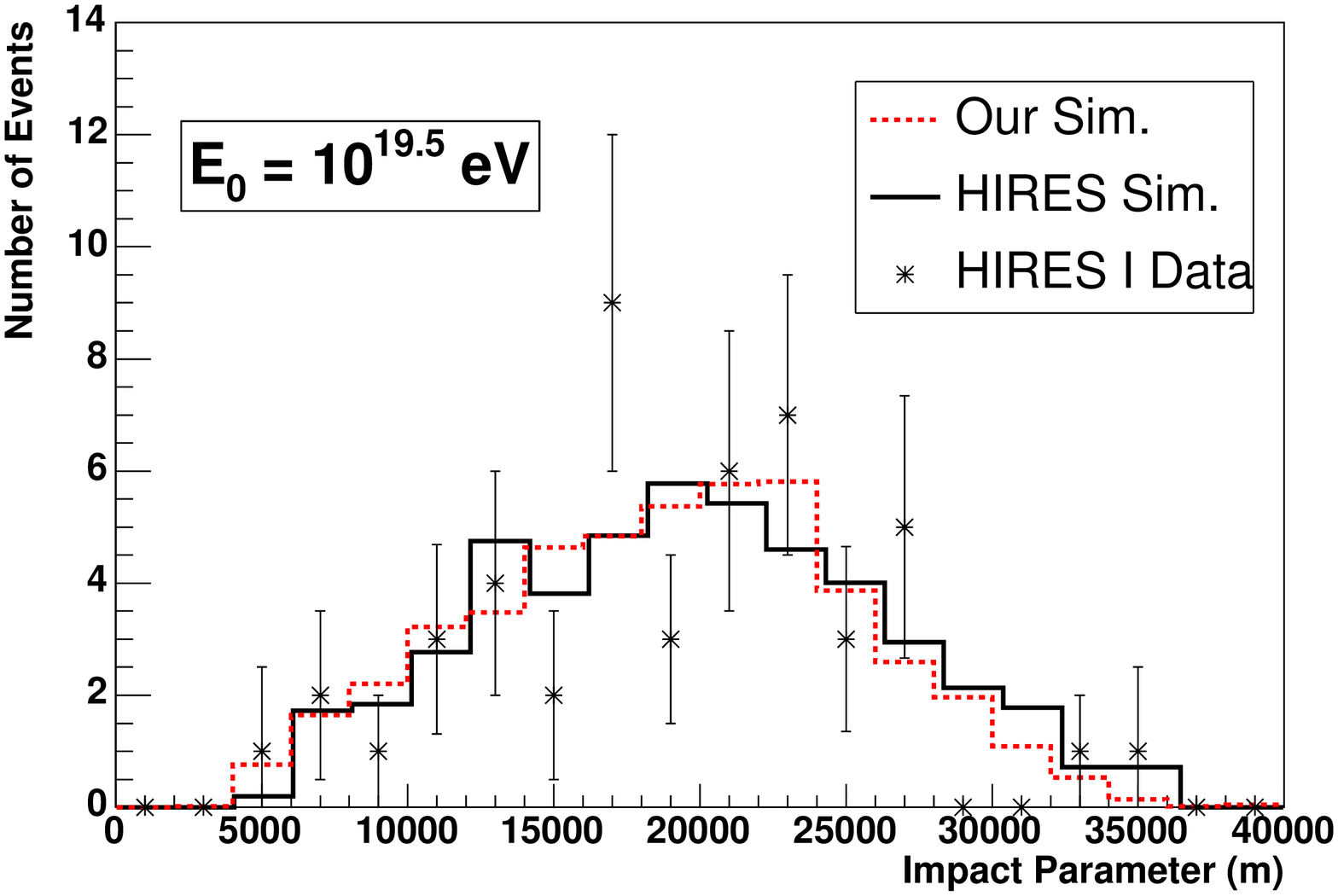}
  \end{center}
  \caption{Distribution of impact parameter for shower initiated with
    energies $10^{19.5}$ eV. We compare our simulation
    with the HiRes-I data and the Hires Collaboration Simulation. The
    $\chi^2$ between the HiRes data and the HiRes simulation is 1.5 and the
    $\chi^2$ between the HiRes data and our simulation is also 1.5. The
   HiRes data and simulation have been extracted from \cite{bib:hires:espectro}.
  }
  \label{fig:rp195}
\end{figure}

Figs.~\ref{fig:rp19} and \ref{fig:rp195} shows a comparison
between our simulation program, the HiRes-I data and the HiRes
Collaboration simulations. The HiRes data and simulation have been
extracted from \cite{bib:hires:espectro}. We have calculated the
impact parameter of the showers detected by the telescopes. The
number of simulated showers was normalized to the number of
detected events.

The good agreement between our simulations and the HiRes data
shows that we were able to reproduce the most important features
of the detector. The $\chi^2$ between our simulation and the data
is 1.7 and 1.5 for $10^{19}$ eV and $10^{19.5}$ eV, respectively,
while the  $\chi^2$ between the HiRes Collaboration simulation and
the data is 1.6 and 1.5 at the same energies.

\section{Reconstruction and Analysis}\label{sec:reconstruction}

The reconstruction of the shower also followed the procedures
described in references
\cite{bib:hires:espectro,bib:hires:espectro:2}. Basically, the
inverse process described in Section \ref{sec:simulation} is
applied in order to reconstruct first the shower longitudinal
profile and subsequently its energy.

The number of photoelectrons measured in each pixel in the
detector is mapped backwards onto (a) the number of photons in the
telescope, (b) the number of photons in the axis of the shower
and, finally, (c) the number of particles in the shower. The
telescope efficiency, atmospheric absorption, fluorescence yield
and missing energies have been considered according to Section
\ref{sec:simulation} and references there in.

A Gaisser-Hillas \cite{bib:gaisser:hillas} function was fitted to
the data according to the specifications in reference
\cite{bib:hires:espectro}. The $X_{\mathrm{max}}$ parameter was
allowed to vary in 35 g/cm$^2$ steps between 680 and 900 g/cm$^2$.
The $X_0$ parameter of the Gaisser-Hillas was fixed to $-60$
g/cm$^2$, according to reference~\cite{bib:hires:espectro:2}

\begin{figure}[t]
  \begin{center}
    \includegraphics[width=10cm]{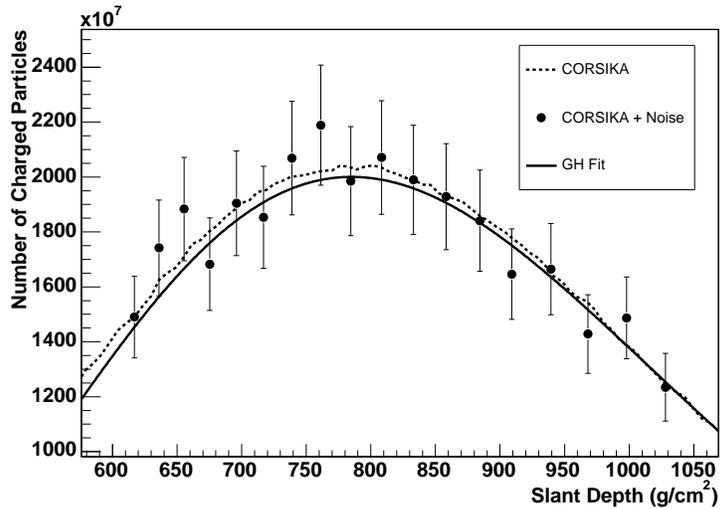}
  \end{center}
   \caption{Longitudinal particles profile simulated by CORSIKA, after
   detector simulation and reconstructed Gaisser-Hillas profile.}
   \label{fig:long}
\end{figure}

Fig.~\ref{fig:long} shows a longitudinal profile of a proton
shower as seen by the HiRes-I telescope according to our
simulation, including noise effects not present in the original
CORSIKA shower, and the longitudinal profile reconstructed from
the detected signal.

Quality cuts are always needed to ensure an accurate
reconstruction. We have required events to satisfy the conditions
listed below as taken from reference \cite{bib:hires:espectro}:

\begin{itemize}
\item Average number of photoelectrons per phototube greater than 25;
\item Angular speed less than $3.33^\circ \ \ \mu s$
\item Track arc-length greater than $8.0^\circ$
\item Depth of first observed point less than 1000 g/cm$^2$
\item Angle of the shower in the plane containing the shower axis and
  the detector greater than $120^\circ$
\end{itemize}

Showers which did not obey those conditions were rejected and
excluded from further analysis. The specific influence of each cut
in the final acceptance of the telescope is the subject of a
detailed study under way at present.

\section{HiRes-I Telescope Aperture}
\label{sec:aperture}

The telescope aperture is essential for the determination of the
spectrum. In order to calculate the HiRes-I telescope aperture we
made a study on the fluctuation of the aperture as a function of
the number of events simulated and the energy of the primary
particle.

We varied the number of showers used to calculate the aperture
from 100 to 5000 by using each CORSIKA shower from 1 to 50 times.
We calculated the aperture 100 times for each total number of
events. Analyzing the distribution of the aperture values we
defined the median, the 68.2\% and 95.4\% confidence levels. The
error is defined as the size of the confidence region divided by
the median aperture.

\begin{figure}[t]
  \begin{center}
    \includegraphics[width=10cm]{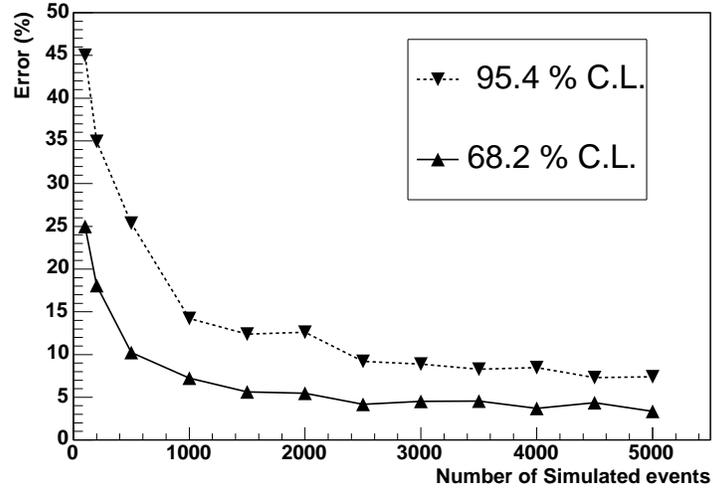}
  \end{center}
   \caption{ Error of the aperture as a function of the
   total number of events simulated.}
   \label{fig:erro:n}
\end{figure}

Fig.~\ref{fig:erro:n} illustrates the  error as a function of the
total number of events simulated for showers initiated by protons
with primary energy $10^{19.5}$ eV. It can be seen that the error
stabilizes around 2500 simulated showers.

\begin{figure}[t]
  \begin{center}
    \includegraphics[width=10cm]{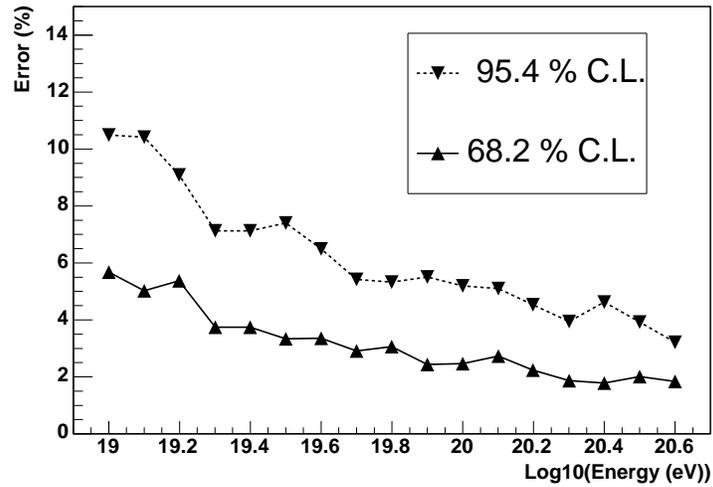}
  \end{center}
   \caption{Error of the aperture as a function of the
   shower energy. We have used 5000 proton shower for each energy.}
   \label{fig:erroEn}
\end{figure}

We repeated the same procedure by calculating the  error as a
function of energy. Fig.~\ref{fig:erroEn} shows the variation of
the error with energy when we used 5000 proton showers in the
simulation. After this analysis, in order to keep the error small
at all energies, we decided to use 5000 showers for the
calculation of the aperture.

\subsection{The dependence of the aperture determination with the
   shower simulation model}
\label{sec:conex}

In order to investigate the influence of the shower simulation
model on the determination of the HiRes-I telescope we used proton
showers simulated with CORSIKA as well as with CONEX.

\begin{figure}[t]
  \begin{center}
    \includegraphics[width=10cm]{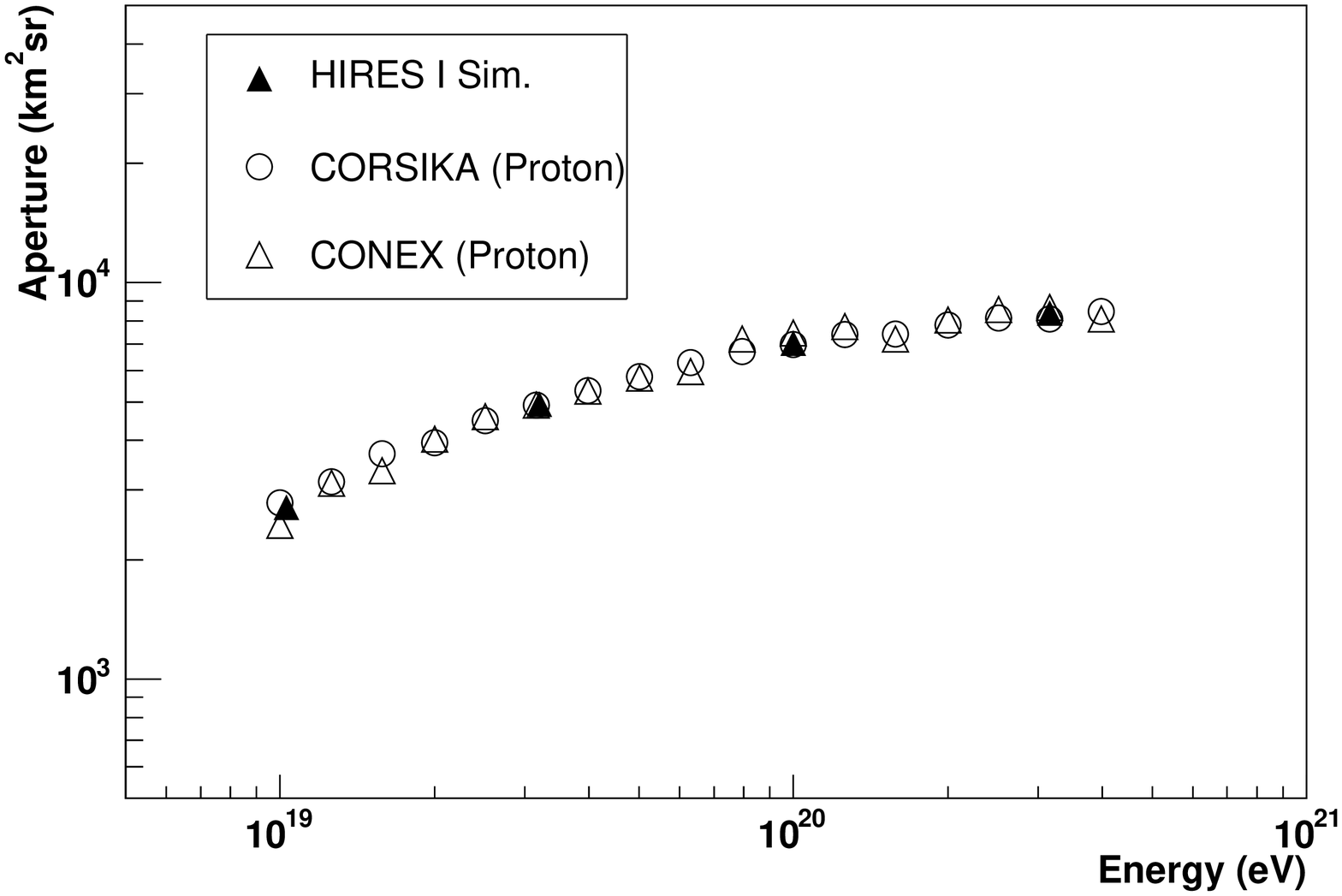}
  \end{center}
   \caption{HiRes-I telescope aperture calculated by the HiRes-I
   Collaboration and published in reference
   \cite{bib:hires:espectro}, calculated with our analysis program
   using shower generated by the CORSIKA and CONEX codes.}
   \label{fig:aperture:conex}
\end{figure}

Fig.~\ref{fig:aperture:conex} shows the aperture for the HiRes-I
telescope for both simulation schemes. First, the figure
illustrates once again that our simulation of the HiRes-I
telescope produces the same results obtained by the HiRes-I
Collaboration, and second, that CORSIKA and CONEX give the same
results concerning aperture calculations. However, despite the
success of this hybrid program (CONEX) in reproducing the same
aperture as given by CORSIKA, we would like to stress that the
aperture calculation is a very indirect procedure to compare
simulation models. More details confrontations between full Monte
Carlo and hybrid simulations have been done as can be seen in
\cite{bib:seneca:corsika,bib:seneca:drescher}.

\begin{figure}[t]
  \begin{center}
    \includegraphics[width=10cm]{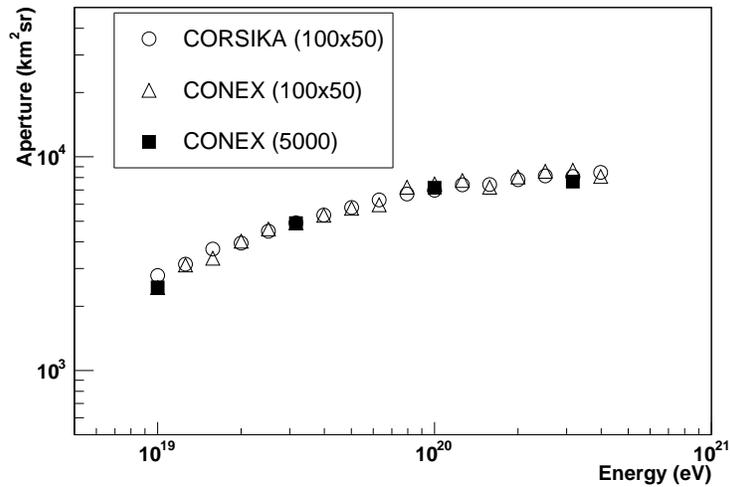}
  \end{center}
   \caption{HiRes-I telescope aperture calculated calculated with our analysis program
     using showers generated by the CORSIKA and CONEX codes in two
     approaches: a) (100x50) simulate 100 different shower and use
     each one 50 times by ramdonly choosing a different geometry and b) (5000)
     simulate 5000 different showers}
   \label{fig:aperture:conex:5000}
\end{figure}

Fig.~\ref{fig:aperture:conex:5000} shows the aperture for the
HiRes-I  telescope calculated with CORSIKA and CONEX. For the
energies of $10^{19}$,  $10^{19.5}$,  $10^{20}$ and  $10^{20.5}$
eV we have simulated a second set of 5000 different showers with
the CONEX program. Fig.~\ref{fig:aperture:conex:5000} shows the
comparison between the two approaches: a) ($100 \times 50$) 100
different showers used 50 times each one by randomly choosing a
different geometry and b) (5000) 5000 different showers. The later
approach was only possible with CONEX because it is much faster
than CORSIKA. No significant difference between the two methods
was seen.

\subsection{The HiRes-I aperture for different primary particles}
\label{sec:aperture:gamma}

We have also investigated the dependence of the aperture on the
nature of the primary particles. According to the article
published by the HiRes Collaboration \cite{bib:hires:espectro}
they have only investigated the aperture of the HiRes-I telescope
for proton and iron nuclei showers and based on this aperture they
have determined the spectrum.

\begin{figure}[t]
  \begin{center}
    \includegraphics[width=10cm]{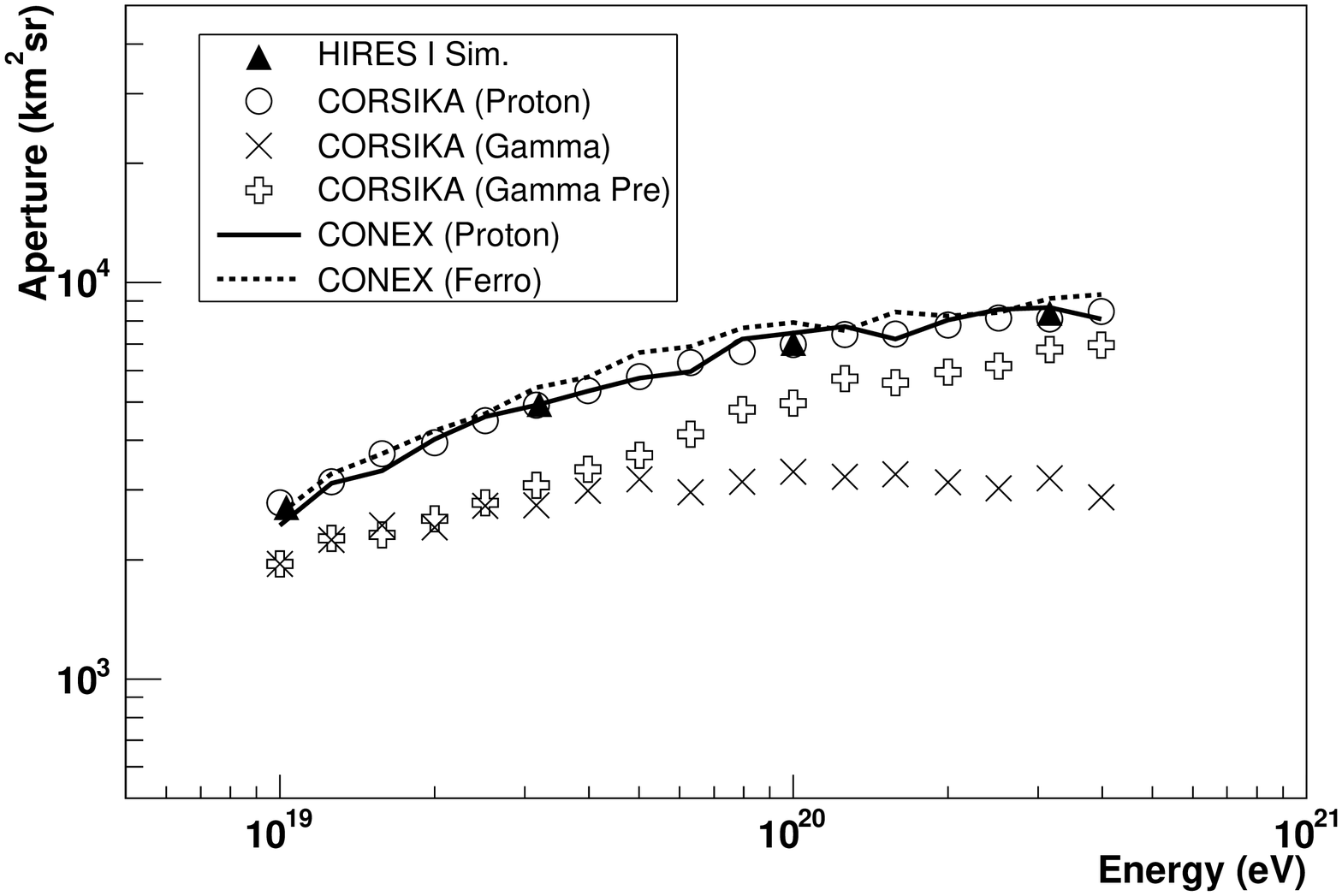}
  \end{center}
   \caption{HiRes-I telescope aperture calculated by the HiRes-I
   Collaboration and published in reference
   \cite{bib:hires:espectro}, calculated with our analysis program
   using showers generated by the CORSIKA and CONEX codes.}
   \label{fig:aperture}
\end{figure}

Fig.~\ref{fig:aperture} shows the aperture for showers initiated
by proton, iron nuclei, gamma without the pre-shower effect and
gamma with the pre shower effect. In all curves, we have used 100
different showers recycled 50 times each as explained in Section
\ref{sec:simulation}.

It is noticeable that the HiRes-I telescope has, within
fluctuations, the same aperture for hadronic showers: proton and
iron nuclei. However, the aperture for gamma showers is well
bellow the hadronic limit.

If the pre shower effect is not taken into account, gamma showers
tends to develop much deeper in the atmosphere, as illustrated in
Fig.~\ref{fig:longitudinal}, when compared to proton showers. The
late development make the gamma showers simulated without the pre
shower effect harder to detect and a great reduction in the telescope
aperture is seen for all energies.

When the pre shower effect is taken into account, gamma shower
with energy above $10^{19.5}$ eV have a considerable probability
of conversion into a pair (more than 5\%) \cite{bib:pre:shower}.
For energies below $10^{19.5}$ eV the probability conversion is
small and no difference is seen in the aperture calculation for
simulation with and without the pre shower effect.

The probability conversion increases very quickly and nears 100\%
between  $10^{20.0}$ and $10^{20.5}$ eV depending on the arrival
direction of the particles in relative to the Earth magnetic
field. The aperture calculation shown in Fig.~\ref{fig:aperture}
illustrates the increase of the conversion probability. Gamma
showers simulated with the pre shower effect evolve from a ``gamma
without preshower profile'' to a ``hadronic profile'' as the
energy varies from $10^{19.5}$ to $10^{20.6}$ eV making the
HiRes-I aperture for gamma showers close to the aperture for the
hadronic shower.

Nevertheless, even at $10^{20.6}$ eV, where the conversion probability
of a gamma in the Earth magnetic field has already reached 100\% the
HiRes-I telescope aperture for gamma showers is smaller than the
aperture for proton or iron nuclei showers.

The effects of the variation of the aperture with energy for different
particle type is taken into account in the next section in order to
calculate the energy spectrum.

\section{Spectrum}
\label{sec:spectrumm}

The fact that the HiRes aperture has been calculated under the
assumption of hadronic primaries, opens the possibility of the
existence of systematic effects for a broad range of cosmic ray
production scenarios.

In particular, it cannot be disregarded at present the possibility
of mixed extragalactic components: a hadronic one, coming from
conservative bottom-up acceleration mechanisms and a harder
component, dominated by photons, originated in more exotic
top-down models.

From the myriad of top-down mechanisms suggested to explain the
observed trans-GZK events \cite{bib:rev:topdown:BhattSigl2000},
the two most promising involve the decay or annihilation of
topological defects and superheavy dark matter
\cite{bib:rev:topdown:Berezinsky,bib:rev:topdown:Kuzmin}. It has
been demonstrated by \cite{bib:topdown:spectra:Aloisio2004} that
the energy spectrum of ultra-high energy cosmic rays generated
from superheavy particles and topological defects is approximately
proportional to $E^{-1.9}$ and that, even if the $\gamma/N$ ratio
is substantially larger than suggested by previous calculations,
photons are dominant above $7-8 \times 10^{19}$ eV. Superheavy
dark matter and several of the topological defects (e.g.,
monopolonia and vortons) cluster strongly in galactic halos and,
therefore, their spectrum at Earth is dominated by the gamma
production spectrum of particles originated in our own halo.
Consequently, the energy spectrum can be approximated by a power
law of spectral index $\sim 2$.

On the other hand, it has been argued by
\cite{bib:topdown:spectra:Aloisio2004} that necklaces, instead of
clustering in the Halo, distribute homogeneously with a rather
small separation, of the order of a few tens of kpc. This means
that the energy spectrum is modified by interactions with the
cosmic microwave background. There is a progressive decrease of
the $\gamma/N$ fraction as energy decreases and protons become
dominant below few $\times 10^{20}$ eV. The resultant spectrum
above $\sim 3 \times 10^{20}$ eV resembles the GZK structure, but
softened and with a rapid recovery.

From the two cases discussed above, clustered and non-clustered
top-down primaries, those that do not cluster are the ones that
best fit the AGASA energy spectrum.

In any case, top-down models can only account for the highest
energy particles, but a more conventional bottom-up model is
needed to explain the rest of the extragalactic spectrum.

As an example, for the bottom-up component, we consider proton
primaries produced by an isotropic distribution of evolving
astrophysical sources.

Indeed, it has been show by \cite{bib:AGN:berezinsky2002} that
regular astrophysical sources, like active galactic nuclei, can
naturally explain the observed energy spectrum between $1$ and
$80$ EeV.

The energy losses of protons in the intergalactic medium are mainly
due to interactions with the comic microwave radiation and, to a
smaller degree, from radio and infrared backgrounds
\cite{bib:book:Berezinsky,bib:thesis:Rachen}. The dominant
contributions come from the expansion of the universe,
electron-positron pair production and photo-pion production.

We characterize energy losses by using the results from
\cite{bib:AGN:berezinsky2002}, which are essentially
indistinguishable from those of \cite{bib:loses:Stanev2000}
calculated with the detailed MC code SOPHIA, and use the
formulation of \cite{bib:spectrum:formalism:BereGrigorieva1988}.
The source luminosity is assumed to have a general redshift
dependence of the form $L=L_{0}(1+z)^{m}$.

\begin{figure}[h]
\begin{center}
  \vspace{5cm}
  \includegraphics[width=10cm,clip=true]{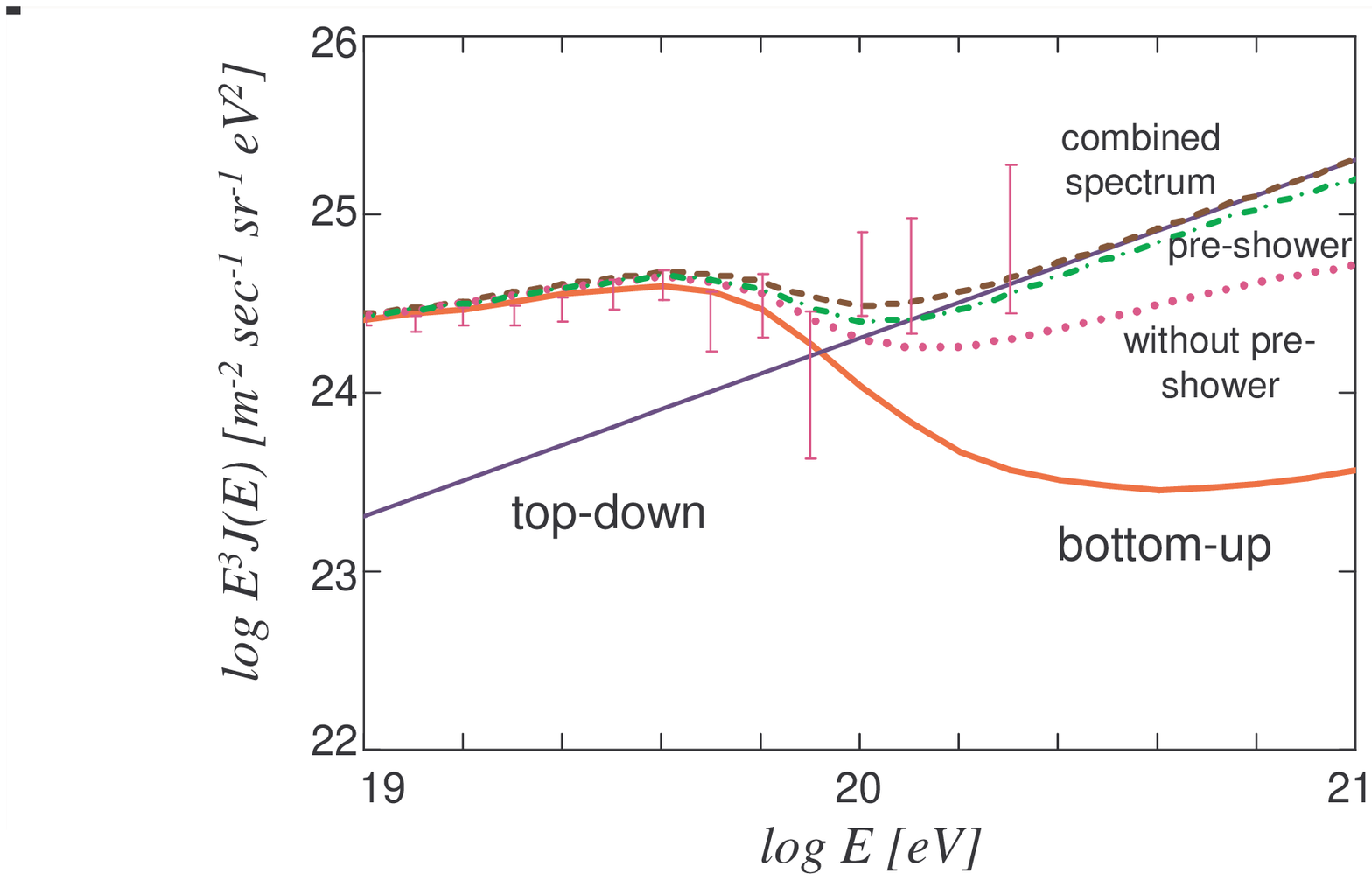}
\caption{\label{fig:espectro} Cosmic ray spectrum as seen by the
  HiRes-I telescope for different primary particles and production
  scenarios (see details in the text), depending whether pre-showering
  is included or not in the calculation of the aperture.}
\end{center}
\end{figure}

As an example, Fig.~\ref{fig:espectro} illustrates a combined
bottom-up/top-down spectrum. The GZK-ed spectrum (lower thick
line) has been numerically calculated using a uniform distribution
of cosmological sources which extends to a maximum redshift
$z_{max}=5$ and whose luminosity evolves with redshift according
to $m=4$. (Actually, since we are only interested on energies
above $10^{19}$ eV, the value of of $z_{max}$ is not important as
long as it is larger than $4$.) Protons were injected at the
sources, with a power law spectrum $dN_{p}/dE \propto E^{-2.5}$)
extending to arbitrary large energies ($E_{max}=\infty$). The
resultant spectrum is normalize to the flux observed by AGASA at
$10^{19}$ eV. The top-down spectrum is assumed dominated by
photons, $dN_{\gamma}/dE \propto E^{-2}$, as expected from the
discussion above. The normalization of the latter spectrum is such
that trans-GZK AGASA data can be fitted by the combined spectrum
(thick dashed line).

The dot dashed curves illustrate the spectrum that HiRes would
infer for the case of a photon component described with and
without pre-showering. It can be seen that in the case without
pre-shower the effect could be severe enough as to make AGASA and
HiRes spectra compatible within quoted uncertainties. The
existence of pre-showers considerably diminishes this effect,
maintaining unaltered the AGASA-HiRes discrepancy. Although only
one realization of the mixed scenario is shown, the conclusions
are not altered by changes in the power index of the generation
spectrum of protons, the evolution rate $m$, maximum injected
energy, $E_{max}$, or reasonable variations in the hardness of the
top-down gamma spectrum.

\section{Conclusion}
\label{sec:conclusion}

The HiRes-I telescope aperture was calculated for different primary
particles (iron nuclei, proton and gamma with and without the pre-shower
effect) and different simulation codes (CORSIKA and CONEX). The
agreement of our calculations with the results published by the HiRes
Collaboration for proton primaries is shown in Fig.~\ref{fig:aperture:conex}.
Same figure shows the agreement between the CONEX and CORSIKA simulation programs.

The recently released program CONEX has shown to be a useful tool for
aperture calculations where a great number of events need to be
simulated. We have tested the effect of using the same longitudinal
profile several time by randomly drawing a new geometry and we concluded
that no significant bias is introduced by this procedure.

We calculated the fluctuations in the aperture determination as a
function of the number of simulated showers and energy. We have shown
that to keep the one sigma fluctuations below 6\% for energies varying
from $10^{19}$ to $10^{20.6}$ eV it is needed to
simulate at least 2500 shower.

The aperture for gamma showers was calculated for the HiRes-I telescope
and has shown to be a factor 1.5 smaller than the aperture for proton
or iron nuclei shower for energies below $10^{19.5}$ eV. For energies above
$10^{19.5}$ eV, the pre-shower effect becomes important and changes
completely the scenario. Gamma showers develop earlier in the
atmosphere due to the interaction with the Earth magnetic field and
the efficiency of detection of the telescopes increases. The resultant
aperture rises getting closer to the aperture for proton
shower. However, even at $10^{20.6}$ eV the telescope aperture for
gammas is a factor 1.2 smaller than the aperture for protons.

The spectrum corresponding to the HiRes-I telescope aperture for gamma
and protons supposing a mixed bottom-up and top-down component is shown
in Fig~\ref{fig:espectro}. The differences in aperture for gamma
and proton showers is not enough to explain the differences between
the AGASA and HiRes spectra. If the pre-shower effect was not taken
into account the AGASA and HiRes spectra would be compatible within the
quoted uncertainties.

\section{Acknowledgments}

This work was supported by the Brazilian science foundations
FAPESP  and CNPq to which we are grateful. Most simulations were
carried out on a Cluster Linux TDI, supported by Laborat\'orio de
Computa\c c\~ao Cient\'{\i}fica Avan\c cada at Universidade de
S\~ao Paulo.

\bibliographystyle{elsart-num}

\bibliography{espectro_ref}

\newpage

\end{document}